\documentclass[preprint,12pt]{elsarticle}
\usepackage[caption = false]{subfig}
\usepackage{array}
\usepackage{amsmath}

\usepackage{amssymb}
\newcounter{bla}

\journal{Computer Physics Communications}

\begin{document}

\begin{frontmatter}

%% Title, authors and addresses

%% use the tnoteref command within \title for footnotes;
%% use the tnotetext command for the associated footnote;
%% use the fnref command within \author or \address for footnotes;
%% use the fntext command for the associated footnote;
%% use the corref command within \author for corresponding author footnotes;
%% use the cortext command for the associated footnote;
%% use the ead command for the email address,
%% and the form \ead[url] for the home page:
%%
%% \title{Title\tnoteref{label1}}
%% \tnotetext[label1]{}
%% \author{Name\corref{cor1}\fnref{label2}}
%% \ead{email address}
%% \ead[url]{home page}
%% \fntext[label2]{}
%% \cortext[cor1]{}
%% \address{Address\fnref{label3}}
%% \fntext[label3]{}

\title{CoronaryHemodynamics: An Automated Simulation Framework for Coronary Artery Hemodynamics Using OpenFOAM}

%% use optional labels to link authors explicitly to addresses:
%% \author[label1,label2]{<author name>}
%% \address[label1]{<address>}
%% \address[label2]{<address>}

\author[a]{Yijin Mao\corref{author}}
\author[a]{Yuwen Zhang}

\cortext[author] {Corresponding author.\\\textit{E-mail address:} maoy@umsystem.edu}
\address[a]{Department of Mechanical and Aerospace Engineering,University of Missouri, Columbia, MO 65211, USA}

\begin{abstract}
%% Text of abstract
CoronaryHemodynamics is a comprehensive simulation package developed on the OpenFOAM platform, designed specifically for coronary artery hemodynamics analysis. The package integrates a complete suite of tools for simulation preparation, including automatic case setup, boundary condition configuration, computational domain meshing, and a CFD solver. It fully supports MPI parallelization, leveraging the native parallel computing capabilities of OpenFOAM. The package implements Windkessel boundary conditions at the aorta outlet and all coronary vessel outlets, with outlet parameters automatically derived from physiological metrics such as heart rate, systolic blood pressure, and myocardial volume. At the aorta inlet, a parabolic flow profile is applied based on the input flow rate waveform. CoronaryHemodynamics supports both steady-state and transient solvers, enabling the simulation of various flow conditions. Flow rate and pressure data are recorded at all boundaries during the simulation, and outputs such as wall shear stress, pressure fields, and velocity fields are automatically stored for detailed post-processing and analysis. By automating critical aspects of the simulation pipeline and integrating physiological boundary conditions, CoronaryHemodynamics offers an efficient and robust framework to study coronary hemodynamics in both research and clinical applications. The package can be found at https://github.com/alundilong/CoronaryHemodynamics.

\end{abstract}


\end{frontmatter}

\section{Introduction}
\label{intro}
The study of coronary artery hemodynamics is critical for understanding cardiovascular physiology and pathology, offering essential insights into the development and progression of diseases such as atherosclerosis \cite{RN12,RN13,RN14,RN15,RN16}. Hemodynamic analysis, particularly through computational approaches, has become an indispensable tool due to its ability to simulate complex blood flow patterns and interactions that are often difficult or impractical to capture experimentally. Computational Fluid Dynamics (CFD) provides a robust framework for modeling blood flow in intricate vascular geometries, enabling researchers to investigate key hemodynamic parameters, such as pressure distributions, wall shear stress, and velocity fields. These parameters are essential for identifying regions susceptible to disease progression and improving treatment planning. CFD has also proven instrumental in clinical applications by facilitating the non-invasive estimation of critical diagnostic indices. Fractional Flow Reserve (FFR) \cite{RN18}, a gold-standard metric for assessing the functional severity of coronary artery stenosis, has been effectively predicted using CFD-based virtual FFR (vFFR) derived from coronary CT angiography (CCTA) \cite{RN8,RN17} . Another widely adopted metric, the Quantitative Flow Ratio (QFR), utilizes angiographic data combined with CFD to evaluate coronary flow and has shown high accuracy and efficiency in diagnosing ischemia-causing lesions \cite{RN9}. These advancements demonstrate the growing clinical relevance of CFD in reducing reliance on invasive procedures while maintaining diagnostic precision. CFD has further applications in predicting long-term cardiovascular outcomes. For instance, CT-FFR metrics such as the maximum pressure drop along coronary arteries have been shown to predict future revascularization needs, demonstrating its utility in guiding both diagnostic and therapeutic decisions \cite{RN10}. Additionally, advanced CFD models, such as closed-loop multi-scale frameworks, integrate physiological and anatomical data to provide highly accurate simulations of coronary flow under realistic conditions \cite{RN11}. \\ 
\indent Over the past decade, numerous open-source software tools have been developed to facilitate patient-specific cardiovascular simulations, addressing both cost and accessibility concerns. Among these tools, SimVascular \cite{RN1} and CRIMSON \cite{RN2} (CardiovasculaR Integrated Modelling and SimulatiON) stand out as leading platforms for cardiovascular modeling. SimVascular provides a comprehensive pipeline for image-based vascular modeling, mesh generation, and blood flow simulation, making it widely used in cardiovascular research and clinical applications. Similarly, CRIMSON offers a powerful and user-friendly software environment for reduced-order and three-dimensional blood flow simulations. CRIMSON supports tasks such as geometric modeling, mesh generation, and post-processing, and it has been widely applied to clinical problems, including pre-operative planning and device optimization. Other notable open-source tools include HeMoLab, which provides a computational environment for human cardiovascular system modeling, and lifex-cfd, a modern C++ based finite element solver tailored for cardiovascular flows under physiological and pathological conditions \cite{RN3,RN4}. These platforms have expanded the capabilities of cardiovascular CFD simulations, allowing researchers to explore patient-specific flow dynamics and improve disease diagnosis and treatment.\\
\indent Among these Open-Source tools, OpenFOAM \cite{RN37,RN38} has gained significant traction as a versatile and widely adopted open-source CFD framework. It provides extensive solvers, robust computational capabilities, and customizable workflows, making it suitable for solving complex fluid dynamics problems, including cardiovascular hemodynamics. Its advanced meshing tools, such as cfMesh \cite{RN6} and pyMeshFOAM \cite{RN5}, enable automatic and efficient meshing of arbitrary and patient-specific geometries, a crucial step in simulating intricate vascular structures like coronary arteries. Given the anatomical variability among patients, efficient meshing tools greatly enhance the applicability and scalability of CFD simulations in biomedical research.\\
\indent Despite the advancements offered by OpenFOAM and other tools, their application to coronary artery hemodynamics remains challenging due to the significant manual effort required for preprocessing, boundary condition setup, and solver configuration. To address these limitations, this study introduces CoronaryHemodynamics, an OpenFOAM-based simulation package specifically designed to streamline the simulation of coronary artery hemodynamics. It integrates pre- and post-processing capabilities, automatic geometry handling (e.g., STL file compatibility), meshing tools such as cfMesh, and physiological boundary condition modeling into a unified and automated pipeline. Its key features include fully automated simulation preparation, implementation of steady-state and transient solvers based on the SIMPLE and PIMPLE algorithms, and compatibility with empirical coronary flow rate formulas as well as circuit-based boundary conditions, including the two- and three-element Windkessel models \cite{RN7}.\\
\indent The CoronaryHemodynamics package is designed to support a broad range of applications, offering detailed analyses of wall shear stress, pressure distributions, and velocity fields. These features are critical for advancing cardiovascular research, such as understanding the progression of atherosclerosis, and for enabling clinical applications, including surgical planning and treatment optimization. By utilizing patient-specific coronary artery geometries extracted from imaging techniques like CCTA, the package bridges the gap between computational flexibility and practical usability. This alignment with the growing emphasis on personalized medicine highlights its potential to deliver meaningful insights for both research and clinical practice.\\
\indent This paper provides a comprehensive overview of the design, implementation, and capabilities of the CoronaryHemodynamics package. It examines the core functionalities, validation processes, and computational performance of the package, while emphasizing its wide-ranging applications in both cardiovascular research and clinical contexts.

\section{Software description}
\label{sd}
\subsection{Overview of the implementation}
\label{sd_overview}

\begin{figure}[h]
\includegraphics[width=\linewidth]{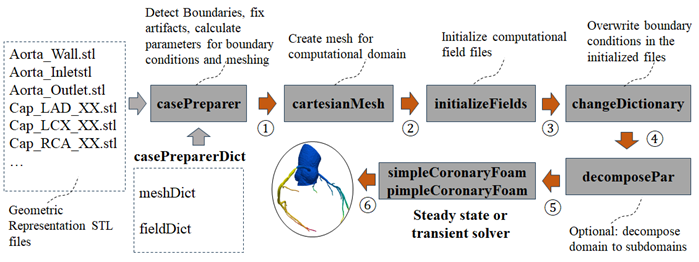}
\centering
\caption{Overview of coronary hemodynamics simulation workflow}
\label{figs:fig1}
\end{figure}
\indent The software comprises six key applications: (1)\textit{casePreparer}, (2)\textit{cartesianMesh}, (3)\textit{initializeFields}, (4)\textit{changeDictionary}, (5)\textit{decomposePar}, (6)\textit{simpleCoronaryFoam} and \textit{pimpleCoronaryFoam}. As depicted in Figure \ref{figs:fig1}, the workflow begins with \textit{casePreparer}, which loads prepared geometric files and uses the accompanying casePreparerDict file to guide the auto-preparation process, including defining boundaries and fixing artifacts. The \textit{cartesianMesh} application, adapted from the third-party software cfMesh, generates a body-fitted cartesian mesh for the computational domain. Next, \textit{initializeFields} creates placeholder pressure and velocity fields for the meshed domain. The \textit{changeDictionary} and \textit{decomposePar} utilities, both native to OpenFOAM, apply boundary conditions generated by \textit{casePreparer} to the initialized pressure and velocity fields. Finally, the solvers \textit{simpleCoronaryFoam} and \textit{pimpleCoronaryFoam} perform steady-state and transient simulations of coronary hemodynamics, respectively. Once the geometric files are prepared and placed in the designated folder, the entire simulation workflow can be executed sequentially from the case directory via the command line, enabling full automation of the process.

\begin{figure}[h]
\includegraphics[width=0.3\textwidth]{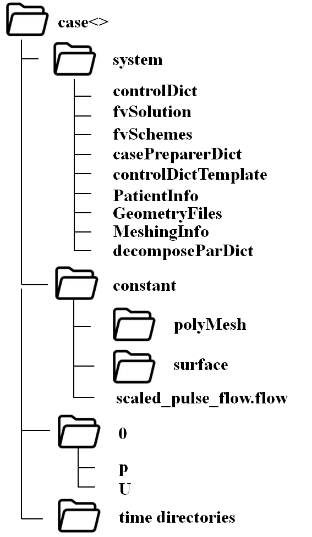}
\centering
\caption{Schematic structure of a coronary hemodynamics case}
\label{figs:fig2}
\end{figure}

\subsection{Case structure}
\label{sd_cs}
\indent Figure \ref{figs:fig2} illustrates the folder structure of a standard OpenFOAM CFD case, comprising three primary directories: 0, constant, and system. The system folder contains standard OpenFOAM files, such as controlDict, fvSolution, and fvSchemes, along with additional files tailored for coronary simulations. The casePreparerDict file specifies key parameters for the \textit{casePreparer} application to process geometry and prepare boundary conditions for the computational domain. The controlDictTemplate file is modified by the application of \textit{casePreparer} depending on whether the simulation is steady-state or transient. The PatientInfo file includes patient-specific physiological parameters, such as heart rate, systolic blood pressure, mean aortic blood pressure, myocardial density, and myocardial volume. The GeometryFiles file specifies the paths to the geometric components of the coronary tree, including the aorta wall and caps for the inlet, outlet, and coronary vessels. The file MeshInfo defines the basic mesh size for discretizing the computational domain. The decomposeParDict file is only required when conducting MPI-based parallel simulations. The constant folder includes the polyMesh and surface directories, along with a flow file of \texttt{scaled\_pulse\_flow.flow}. The surface directory stores geometry files, while the \texttt{scaled\_pulse\_flow.flow} file provides the waveform profile for the aorta inflow. The polyMesh directory holds the mesh data generated after running the \textit{cartesianMesh} application. The 0 directory contains the initial and boundary conditions for pressure and velocity fields, stored in the standard OpenFOAM format.

\subsection{Theoretical Models}
\label{tm}
\subsubsection{Governing Equations}
\label{tm_ge}
\indent The simulation framework adopts the governing equations for incompressible Newtonian fluid dynamics, assuming constant fluid density and viscosity. This approach is particularly relevant for simulating blood flow, where the incompressibility assumption is valid due to the negligible compressibility of blood under physiological conditions. The governing equations consist of the continuity equation and the momentum conservation equation, which together describe the fluid motion.\\
\indent The continuity equation, expressed as,
\begin{equation}
\nabla \cdot (\rho \mathbf{U}) = 0
\end{equation}
ensures mass conservation within the flow domain. Here, $\rho$ represents the fluid density, and $\mathbf{U}$ denotes the velocity field. For an incompressible fluid, $\rho$ is constant, reducing the equation to $\nabla \cdot \mathbf{U} = 0$, which implies that the velocity field has zero divergence.
\\
\indent The momentum conservation equation is given by,
\begin{equation}
\frac{\partial\rho\mathbf{U}}{\partial t}+\nabla\left(\rho\mathbf{U}\otimes\mathbf{U}\right)=-\nabla p+\\{\mathbf{R}}
\end{equation}
where $p$ refers to the pressure field, and $\mathbf{R}$ denotes the stress tensor, which accounts for the viscous forces acting within the fluid. For a Newtonian fluid, the stress tensor $\mathbf{R}$ is related to the velocity gradient through the fluid's viscosity. These equations form the mathematical foundation for computational simulations of fluid dynamics in coronary arteries. They are solved numerically using finite volume methods implemented in OpenFOAM. 
\begin{figure}[h]
\includegraphics[width=0.3\textwidth]{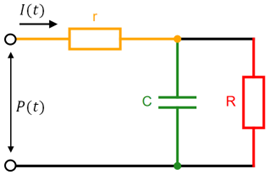}
\centering
\caption{Circuit model of Three-Element Windkessel Model}
\label{figs:fig3}
\end{figure}
\subsubsection{Boundary Conditions}

\begin{figure}[h]
\includegraphics[width=0.3\textwidth]{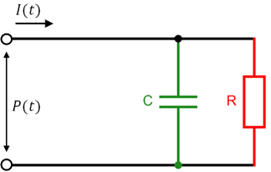}
\centering
\caption{Circuit model of Two-Element Windkessel Model}
\label{figs:fig4}
\end{figure}

\indent The boundary conditions for the simulation are based on the Windkessel model, which is a well-established framework in hemodynamic studies for representing the relationship between pressure and flow in blood vessels. Two variants of the Windkessel model are employed: the three-element Windkessel model and the two-element Windkessel model. These models are used to provide physiologically reasonable representations of the vascular impedance and flow dynamics at the outlet boundaries.\\
\indent As shown in Figure \ref{figs:fig3}, the three-element Windkessel model \cite{RN21,RN25,RN26} incorporates resistance ($R$), compliance ($C$), and an additional proximal resistance ($r$) to account for both the resistance of the vascular system and the compliance of the arterial walls. The governing equation for the three-element model is expressed as,
\begin{equation}
\left(1+\frac{r}{R}\right)I\left(t\right)+Cr\frac{dI\left(t\right)}{dt}=\ \frac{P\left(t\right)}{R}+C\frac{dP\left(t\right)}{dt}
\end{equation}
where $P(t)$ represents the pressure at the boundary, $I(t)$ is the flow rate, $R$ is the distal resistance, $r$ is the proximal resistance, and $C$ is the compliance of the arterial walls. This model provides a more detailed approximation of the dynamic vascular response by including the proximal resistance component.
\\
\indent As given in Figure \ref{figs:fig4}, the two-element Windkessel model is a simplified representation that excludes the proximal resistance, focusing solely on the compliance and distal resistance of the vasculature. The governing equation for the two-element model is as follows,
\begin{equation}
I\left(t\right)=\ \frac{P\left(t\right)}{R}+C\frac{dP\left(t\right)}{dt}
\end{equation}
In this model, the flow rate $I(t)$ depends on the pressure $P(t)$, distal resistance $R$, and arterial compliance $C$. While simpler, this model is still widely used in cases where the proximal resistance is negligible or where computational efficiency is a priority \cite{RN20,RN22,RN23}.\\
\indent These boundary conditions are crucial for simulating the interaction between blood flow and the vascular system. The Windkessel models provide the necessary impedance at the outlets, ensuring realistic pressure and flow waveforms that closely mimic physiological conditions. The inclusion of compliance captures the elastic behavior of arterial walls, while the resistive components represent the opposition to blood flow due to vascular resistance. These boundary conditions enable the computational framework to deliver robust and physiologically reasonable hemodynamic.
\begin{figure}[h]
\includegraphics[width=\textwidth]{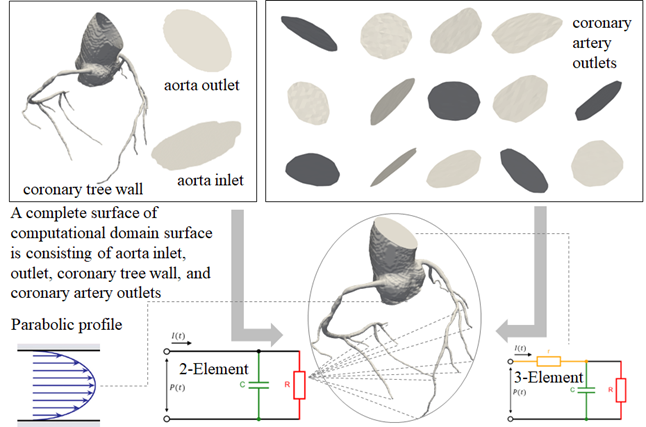}
\centering
\caption{Decomposition of the computational domain and implementation of boundary conditions}
\label{figs:fig5}
\end{figure}

\subsubsection{Physical Parameters for Windkessel Models}
\label{tm_ppwm}
The parameters for the Windkessel models used in the simulation are derived using physiological data and empirical relationships to ensure an accurate representation of coronary hemodynamics. The total coronary flow \cite{RN29}, $Q_{cor}$ is estimated using an empirical formula that incorporates key physiological parameters, as given in Equation 5, including heart rate (HR), systolic blood pressure (SBP), and myocardial mass ($M=\rho_MV_M$),
\begin{equation}
Q_{cor}=1.5\times\left[0.08\times\left(0.0007HR\ \times\ SBP\right)-0.4\right]\frac{M}{60}
\end{equation}
\indent This formula provides the mean flow through the coronary arteries, serving as the basis for determining the total resistance applied to the coronary system. The total resistance, $R_{total}$, is computed using the relationship below,
\begin{equation}
R_{total}=1333.33\frac{MAP}{Q_{cor}}\times{10}^5
\end{equation}
where $MAP$ refers to the mean aortic pressure, which drives blood flow through the coronary system. The resistances at individual coronary outlets are then distributed using Murray’s Law \cite{RN28,RN30}, which relates the resistance at each outlet to the vessel’s cross-sectional area $A_i$. The resistance at each outlet $R_i$ is given by,
\begin{equation}
R_i=R_{total}\frac{\sum_{i=0}^{N}A_i^{1.3}}{A_i^{1.3}}
\end{equation}
\begin{equation}
R_{i,a}={r_{i,a}R}_i
\end{equation}
\begin{equation}
R_{i,amicro}={r_{i,amicro}R}_i
\end{equation}
\begin{equation}
R_{i,v}={(1-r_{i,a}-r_{i,amicro})R}_i
\end{equation}
where $A_i^{1.3}$ represents the weighted contribution of each outlet based on its cross-sectional area \cite{RN28}. To further capture regional vascular behavior, resistances are subdivided into three components \cite{RN32,RN33}: large artery resistance ($R_{i,a}={r_{i,a}R}_i$), microvascular resistance ($R_{i,amicro}={r_{i,amicro}R}_i$), and venous resistance ($R_{i,v}={(1-r_{i,a}-r_{i,amicro})R}_i$).

\indent Similarly, arterial compliance is distributed across the coronary tree using the same scaling principle. Compliance is treated separately for the left and right coronary arteries. The compliance at each outlet, $C_{L,i}$ for the left coronary, and $C_{R,i}$ for the right coronary, is calculated as:
\begin{equation}
C_{L,i}=C_{L,total}\frac{A_i^{1.3}}{\sum_{i=0}^{N_{Left}}A_i^{1.3}}
\end{equation}

\begin{equation}
C_{L,i,a}={r_{L,i,a}C}_{L,i}
\end{equation}

\begin{equation}
C_{L,i,im}=(1-r_{L,i,a})C_{L,i}
\end{equation}

\begin{equation}
C_{R,i}=C_{R,total}\frac{A_i^{1.3}}{\sum_{i=0}^{N_{Right}}A_i^{1.3}}
\end{equation}

\begin{equation}
C_{R,i,a}={r_{R,i,a}C}_{R,i}
\end{equation}

\begin{equation}
C_{R,i,im}=(1-r_{R,i,a})C_{R,i}
\end{equation}
where $C_{L,total}$ and $C_{R,total}$ represent the total compliance of the left and right coronary arteries, respectively. These are further divided into two subcomponents: large arterial compliance ($C_{L,i,a}={r_{L,i,a}C}_{L,i}$) and microvascular compliance ($C_{L,i,im}=(1-r_{L,i,a})C_{L,i}$).
This parameterization framework ensures that the distribution of resistance and compliance is physiologically reasonable, capturing the heterogeneous nature of coronary flow dynamics.

\begin{table}[h!]
\centering
\renewcommand{\arraystretch}{1.5}
\scriptsize
\begin{tabular}{|>{\raggedright\arraybackslash}m{4.5cm}|>{\raggedright\arraybackslash}m{2.5cm}|>{\centering\arraybackslash}m{3cm}|>{\centering\arraybackslash}m{2cm}|}
\hline
\textbf{Key entries in \texttt{casePrepareDict}} & \textbf{Math. symbol} & \textbf{Default value} & \textbf{Units} \\ \hline
\texttt{coronaryLeftCapactanceTotal} & $C_{L,\text{total}}$ & $3.6 \times 10^{-9}$ & kg/m\textsuperscript{2}/Pa \\ \hline
\texttt{coronaryRightCapactanceTotal} & $C_{R,\text{total}}$ & $2.5 \times 10^{-9}$ & kg/m\textsuperscript{2}/Pa \\ \hline
\texttt{CaCcorRatio} & $r_{R,i,a}$ and $r_{L,i,a}$ & $0.11$ & 1 \\ \hline
\texttt{RaRcorRatio} & $r_{i,a}$ & $0.32$ & 1 \\ \hline
\texttt{RamicroRcorRatio} & $r_{i,\text{amicro}}$ & $0.52$ & 1 \\ \hline
\texttt{windKesselAortaC} & $C$ & $1.0 \times 10^{-9}$ & kg/m\textsuperscript{2}/Pa \\ \hline
$MAP$ & $MAP$ & NA & mmHg \\ \hline
$HR$ & $HR$ & NA & Beat/min \\ \hline
$SBP$ & $SBP$ & NA & mmHg \\ \hline
\texttt{MyocardialDensity} & $\rho_M$ & NA & g/mL \\ \hline
\texttt{MyocardialVolume} & $V_M$ & NA & mL \\ \hline
\end{tabular}
\caption{Key physiological parameters for the coronary hemodynamics simulation.}
\label{tab:table1}
\end{table}

\subsection{Application of \textit{casePreparer}}
\label{app_case}
\indent The \textit{casePreparer} application is the first and most critical step in the automated CoronaryHemodynamics workflow, as depicted in Figure \ref{figs:fig5}. It prepares the computational domain for coronary artery hemodynamics simulations by automating geometry processing, boundary condition assignment, and meshing configuration. By organizing and parameterizing raw anatomical inputs, \textit{casePreparer} creates a simulation-ready domain that seamlessly feeds into subsequent preprocessing stages.\\
\indent The application begins with input geometric files in STL format, representing the primary boundaries of the computational domain: the aorta inlet, aorta outlet, coronary artery outlets, and the coronary tree wall. These surfaces, extracted from medical imaging techniques such as Coronary Computed Tomography Angiography, are critical for defining the regions where blood enters, exits, and interacts with vessel walls. To maintain clarity and consistency, a standardized naming convention is employed for all geometric files. Files prefixed with Cap are designated as inlet/outlet boundaries. The second keyword, separated by an underscore, specifies the main vessel name. For example, \texttt{Cap\_Aorta} refers to the aorta inlet or outlet, while \texttt{Cap\_LAD} represents the outlet of the left anterior descending (LAD) coronary artery. When required, a third keyword specifies a lower-level branch of the main vessel; for instance, \texttt{Cap\_LAD\_D1} identifies the first diagonal branch of the LAD artery. This systematic approach ensures accurate identification, distinction, and parameterization of coronary vessels, which is particularly essential for realistic boundary condition assignment and downstream analyses.\\
\indent The \textit{casePreparer} processes these inputs by assigning appropriate boundary conditions to each surface. At the aorta inlet, a parabolic velocity profile is applied to simulate the pulsatile flow conditions observed in physiological systems. For the coronary artery outlets, the boundary conditions are modeled using the Windkessel framework, which represents vascular impedance and compliance. The two-element Windkessel model incorporates resistance and compliance to simulate simplified outlet dynamics efficiently. In contrast, the three-element Windkessel model extends this by including proximal resistance, capturing upstream effects on flow and pressure more comprehensively. This dual-model capability enables flexibility in balancing computational efficiency and physiological accuracy.
In addition to boundary condition assignment, \textit{casePreparer} calculates key meshing parameters and prepares instructions for downstream meshing processes. The tool employs cfMesh to generate a structured Cartesian mesh tailored to the coronary geometries. Special care is taken to address geometric imperfections, such as gaps or artifacts in the STL files, ensuring a smooth and artifact-free computational surface. These corrections enhance mesh quality and solver stability, which are critical for high-fidelity simulations.\\
\indent The outputs of \textit{casePreparer} include fully parameterized boundary conditions, meshing specifications, and computational field configurations. These outputs are systematically organized and serve as inputs to subsequent preprocessing applications, as shown in Figure \ref{figs:fig5}. The \textit{cartesianMesh} application along with cfMesh utilizes the generated parameters to create a computational mesh, while the \textit{initializeFields} application initializes pressure and velocity fields across the domain. This structured workflow eliminates the need for manual intervention, ensuring consistency and reproducibility across simulations.\\
\indent By automating boundary condition assignment, meshing preparation, and geometric parameterization, \textit{casePreparer} significantly reduces manual effort while ensuring accuracy and clarity in the simulation setup. Its integration of a standardized file-naming convention allows precise identification of coronary vessels, distinguishing between major outlets and lower-level branches within the left and right coronary trees. This systematic preparation ensures that the computational domain is physiologically reasonable and optimized for numerical solvers, forming the foundation for high-fidelity coronary hemodynamics simulations.\\
\indent The physiological parameters loaded via \textit{casePreparer} are summarized in Table \ref{tab:table1}. They play a critical role in defining the boundary conditions for the Windkessel models applied at the coronary artery outlets. These parameters ensure the reasonable representation of vascular resistance and compliance, both of which are essential for capturing realistic hemodynamic responses. The total capacitance for the left and right coronary arteries ($C_{L,total}$ and $C_{R,total}$) governs the compliance of the arterial walls, allowing for elastic behavior under varying pressure conditions. Proximal resistance ratios ($r_{i,a}$) and microvascular resistance ratios ($r_{i,amicro}$) are employed to distribute resistance across the vascular network, with the compliance parameter $windKesselAortaC$ further ensuring appropriate impedance at the aorta outlet. Patient-specific metrics such as mean aortic pressure ($MAP$), heart rate ($HR$), and systolic blood pressure ($SBP$) influence the calculation of flow rates and resistances, which are integral to both the two-element and three-element Windkessel frameworks. By incorporating these parameters, the models achieve physiologically reasonable pressure-flow relationships at the coronary outlets, thereby enhancing the reliability of the simulation results.

\section{Validation}
\label{val}

\begin{figure}
\subfloat[computational mesh for the pipe flow]{\includegraphics[width = 0.5\linewidth]{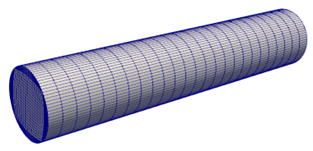}} 
\subfloat[Flow rate profile at inlet]{\includegraphics[width = 0.5\linewidth]{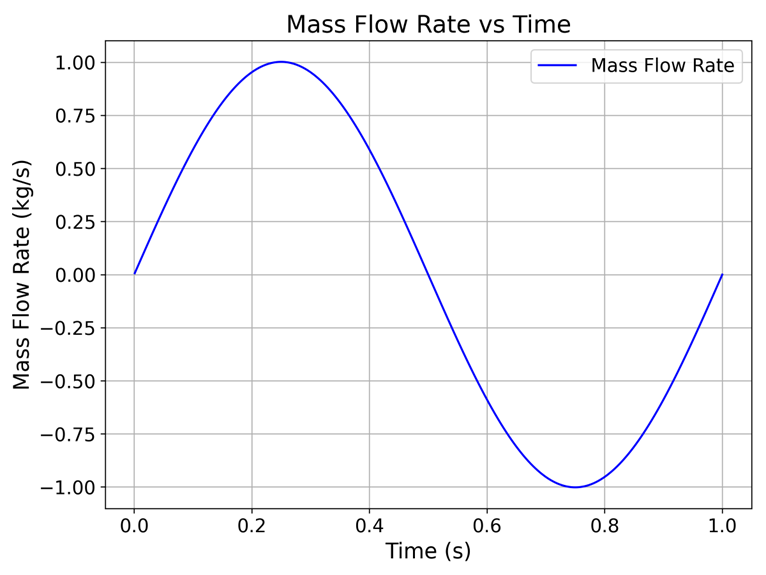}}
\caption{Mesh configuration and flow rate profile imposed to the test case}
\label{figs:fig6}
\end{figure}

\indent To verify the accuracy and robustness of the numerical implementation of the Windkessel model in the CoronaryHemodynamics package, two pipe flow cases are designed using identical sinusoidal inflow conditions but with different outlet boundary conditions. Since the Windkessel model is zero-dimensional, solved only at the boundaries, any simple geometry is sufficient for validation purposes. Here, a straight cylindrical pipe was chosen as the computational domain to ensure simplicity and eliminate geometric complexities. \\
\begin{figure}[h]
\includegraphics[width=0.5\textwidth]{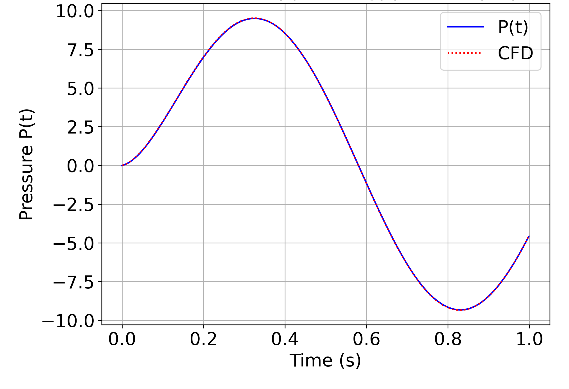}
\centering
\caption{Comparison of pressure variations between the analytical solution and CFD simulation at the outlet with the Three-Element Windkessel boundary condition}
\label{figs:fig7}
\end{figure}
\indent The computational domain for the validation is illustrated in Figure \ref{figs:fig6}(a), consisting of a cylindrical geometry with a length of 5.0 m and a radius of 0.5 m. The inflow rate, $Q(t)$, is defined mathematically as $Q(t)=Asin(\omega t)$, where $A$ is set to 1 and $\omega$ is $2\pi$. This formulation ensures a well-defined and periodic flow condition for precise comparison with analytical results. 
\begin{figure}[h]
\includegraphics[width=0.5\textwidth]{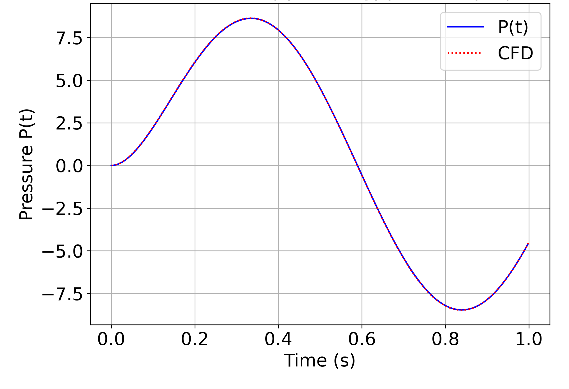}
\centering
\caption{Comparison of pressure variations between the analytical solution and CFD simulation at the outlet with the Two-Element Windkessel boundary condition}
\label{figs:fig8}
\end{figure}
\subsection{Pipeflow-I (Three-Element model)}

\indent For the three-element Windkessel model, the parameters were set as $r=1$, $R=10$, and $C=0.01$. The computational solution was obtained using the native \textit{pimpleFoam} solver, while the analytical solution was generated through the SciPy \cite{RN34} initial value problem (IVP) solver. The comparison of outlet pressure between the CFD simulation and the analytical solution is shown in Figure \ref{figs:fig7}. The pressure profile at the outlet, derived from both methods, matches precisely, confirming the consistency of the CFD implementation. 

\subsection{Pipeflow-II  (Two-Element model)}
\label{val_pipe2}
\indent For the two-element Windkessel model, the parameters were set as $R=10$ and $C=0.01$. Similar to the three-element case, the CFD solution was obtained using the \textit{pimpleFoam} solver, and the analytical solution was computed via the SciPy IVP solver. The pressure profiles obtained from the numerical and analytical approaches are shown in Figure \ref{figs:fig8}. As with the three-element model, the pressure profiles for the two-element model exhibit an exact match between the CFD and analytical solutions. \\
\indent The validation results demonstrate that the CoronaryHemodynamics package provides accurate numerical solutions for both the two-element and three-element Windkessel models. 

\section{Demonstration}
\label{demo}
\indent The following section presents two demonstration cases to highlight the capabilities of the CoronaryHemodynamics package in simulating coronary artery hemodynamics. The first case involves steady-state simulations using a real coronary artery geometry \cite{RN35,RN36}, applying the three-element Windkessel model at the aorta outlet and the two-element Windkessel model at all coronary outlets. This case demonstrates the framework's ability to reproduce expected flow and pressure behaviors under physiological boundary conditions. The second case demonstrates transient simulations over multiple cardiac (10) cycles, showcasing the solver’s ability to capture pulsatile flow dynamics and velocity field variations within anatomically realistic coronary geometries. These examples emphasize the robustness, accuracy, and applicability of the package for analyzing both steady and transient coronary hemodynamics. 

\subsection{CoronaryFlow-steady state simulation}
\label{demo_ss}
\begin{figure}
\subfloat[WSS]{\includegraphics[width = 0.35\linewidth]{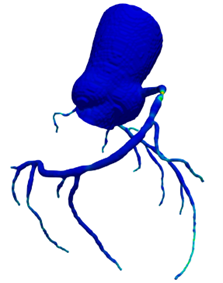}} 
\subfloat[Pressure]{\includegraphics[width = 0.33\linewidth]{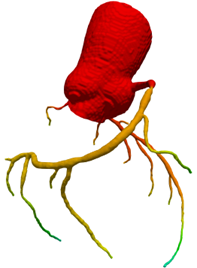}}
\subfloat[Velocity]{\includegraphics[width = 0.33\linewidth]{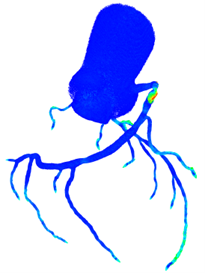}}
\caption{Colormap of wall shear stress, pressure and velocity when steady state simulation converged}
\label{figs:fig9}
\end{figure}
\indent The following demonstration case employs geometric files extracted from a real coronary artery tree to simulate blood flow under steady-state conditions. The boundary conditions are defined using the three-element Windkessel model at the aorta outlet and the two-element Windkessel model at all coronary artery outlets. Under steady-state conditions, the pressure at each outlet is expected to satisfy the relationship $P=Q\cdot R$, where $Q$ is the flow rate and $R$ is the resistance. The computational domain consists of a total mesh size of 300,419 cells. The simulation was executed on a laptop equipped with a $12^{th}$ Gen Intel(R) Core(TM) i7-12700H processor using a single CPU core. The total computational time was 31 seconds.\\
\begin{figure}
\subfloat[Flow rate variation at inlet and outlets along with steady state simulation]{\includegraphics[width = \linewidth]{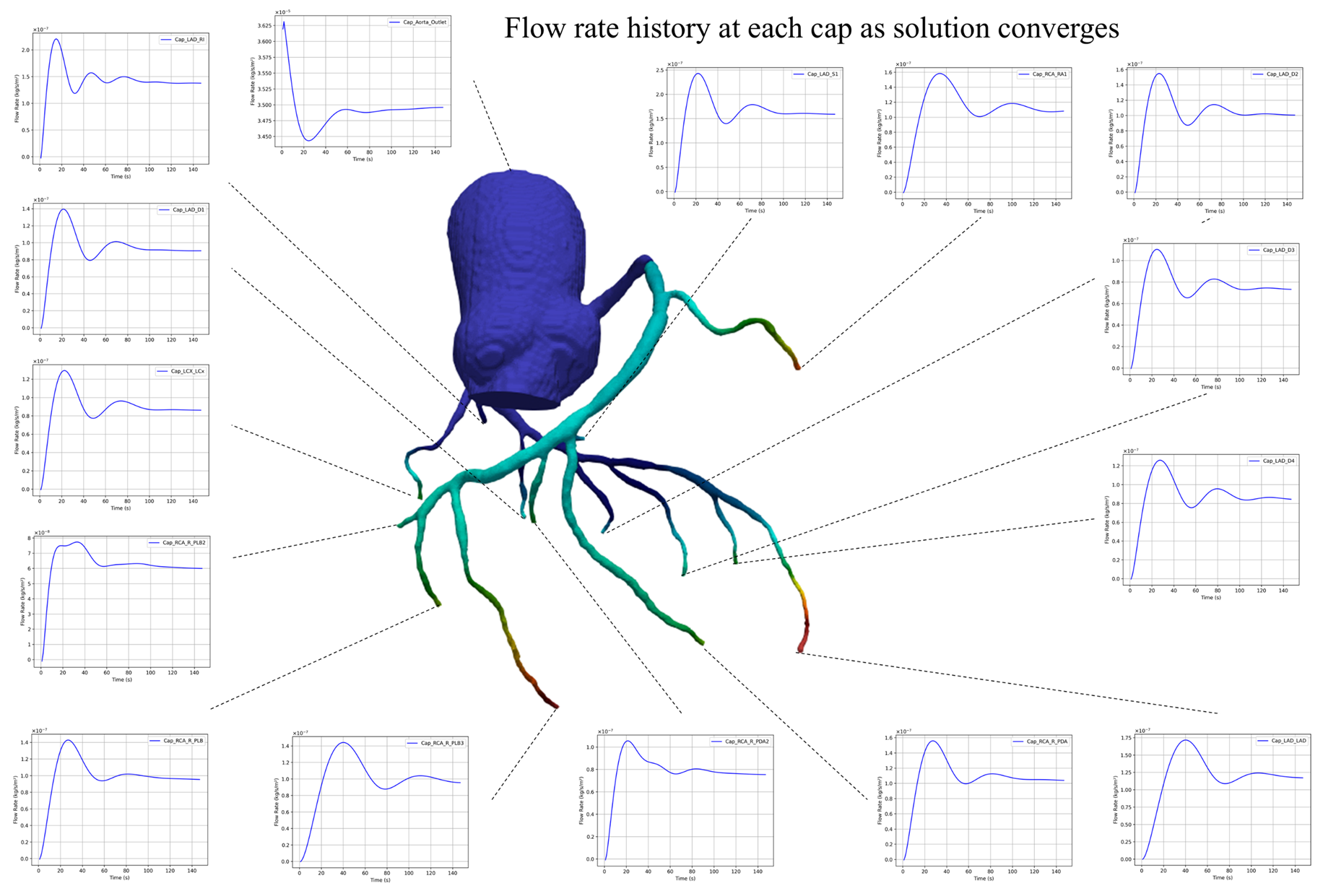}} \\
\subfloat[Pressure variation at inlet and outlets along with steady state simulation]{\includegraphics[width = \linewidth]{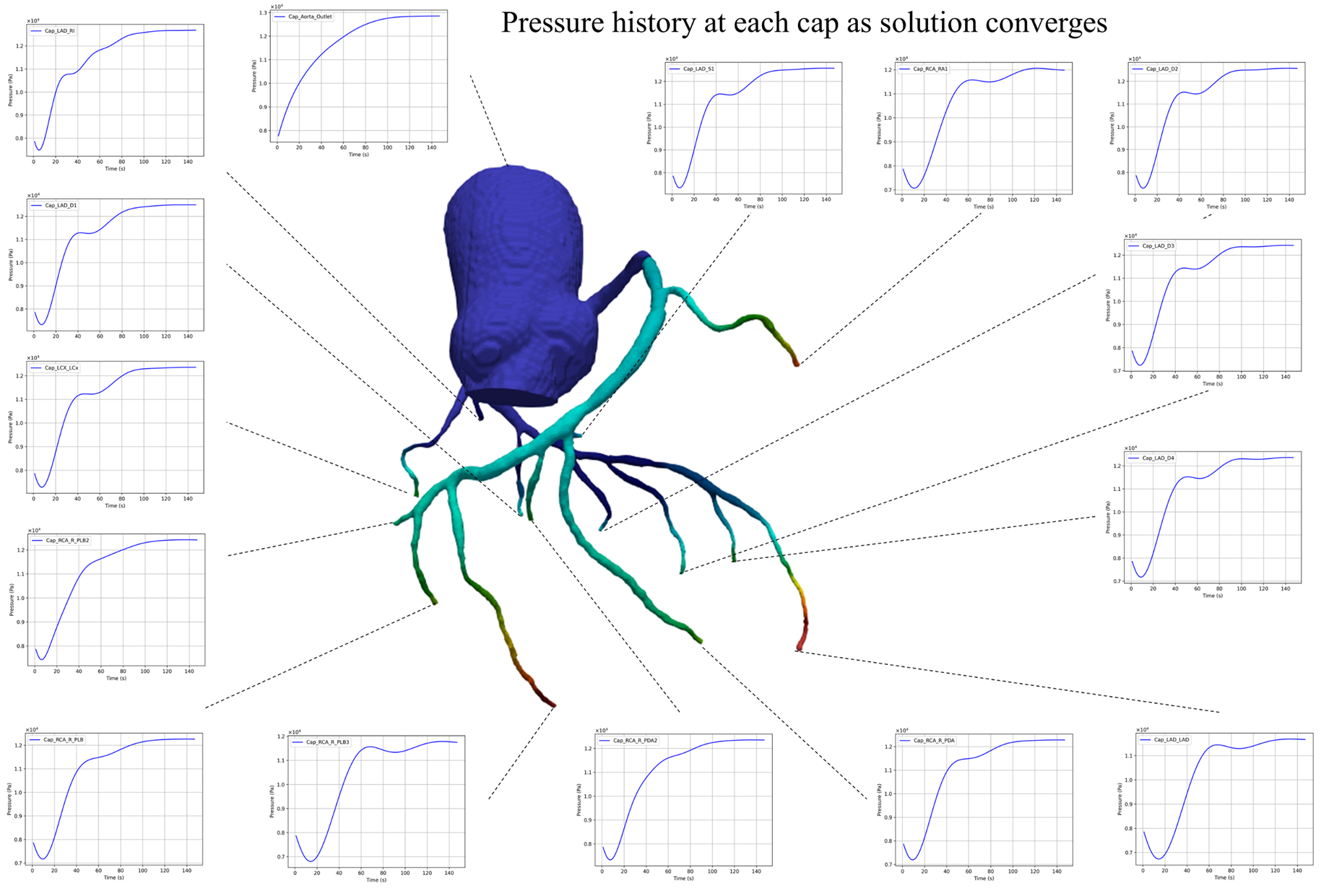}}
\caption{Variation of flow rate and pressure at inlet and outlet boundaries}
\label{figs:fig10}
\end{figure}
\indent Figure \ref{figs:fig9} presents the final distributions of wall shear stress (WSS), pressure, and velocity across the computational domain, obtained using the steady-state solver \textit{simpleCoronaryFoam}. These results provide a comprehensive depiction of the hemodynamic environment within the coronary arteries, offering valuable insights into the flow behavior and pressure gradients. Figure \ref{figs:fig10} illustrates the variations of flow rate and pressure at the inlet and outlets as the simulation progresses toward steady-state conditions. Initially, the flow rate and pressure curves exhibit noticeable oscillations as the numerical solution adjusts to the imposed boundary conditions. Over time, these oscillations gradually decay as the simulation stabilizes, and the curves converge to constant values. Specifically, the flow rate at each outlet stabilizes to a near-constant value, consistent with the prescribed boundary condition, while the pressure at each outlet converges to values governed by the Windkessel model. The converged pressure values align closely with the theoretical relationship $P=Q\cdot R$, further validating the numerical implementation of the Windkessel boundary conditions.\\
\begin{table}[h!]
\centering
\renewcommand{\arraystretch}{1.5}
\scriptsize
\begin{tabular}{|>{\raggedright\arraybackslash}m{2.5cm}|>{\raggedright\arraybackslash}m{2.5cm}|>{\centering\arraybackslash}m{2.5cm}|>{\centering\arraybackslash}m{2cm}|>{\centering\arraybackslash}m{2cm}|}
\hline
\textbf{Cap Name} & \textbf{Flow rate $I$ (kg/s/m\textsuperscript{2})} & \textbf{Total Resistance $R_{\text{total}}$ (Pa m\textsuperscript{2}/kg)} & \textbf{Pressure $P$ (Pa)} & \textbf{$\frac{|P - IR|}{P}$} \\ \hline
Cap\_LAD\_D1 & $9.048 \times 10^{-8}$ & $1.382 \times 10^{11}$ & $1.250 \times 10^{4}$ & $4.99 \times 10^{-4}$ \\ \hline
Cap\_LAD\_D2 & $1.006 \times 10^{-7}$ & $1.246 \times 10^{11}$ & $1.256 \times 10^{4}$ & $2.96 \times 10^{-3}$ \\ \hline
Cap\_LAD\_D3 & $7.316 \times 10^{-8}$ & $1.691 \times 10^{11}$ & $1.242 \times 10^{4}$ & $4.28 \times 10^{-3}$ \\ \hline
Cap\_LAD\_D4 & $8.442 \times 10^{-8}$ & $1.454 \times 10^{11}$ & $1.236 \times 10^{4}$ & $6.74 \times 10^{-3}$ \\ \hline
Cap\_LAD\_LAD & $1.171 \times 10^{-7}$ & $9.872 \times 10^{10}$ & $1.167 \times 10^{4}$ & $9.01 \times 10^{-3}$ \\ \hline
Cap\_LAD\_RI & $1.375 \times 10^{-7}$ & $9.231 \times 10^{10}$ & $1.268 \times 10^{4}$ & $1.31 \times 10^{-3}$ \\ \hline
Cap\_LAD\_S1 & $1.589 \times 10^{-7}$ & $7.902 \times 10^{10}$ & $1.258 \times 10^{4}$ & $1.69 \times 10^{-3}$ \\ \hline
Cap\_LCX\_Lcx & $8.612 \times 10^{-8}$ & $1.434 \times 10^{11}$ & $1.236 \times 10^{4}$ & $2.73 \times 10^{-4}$ \\ \hline
Cap\_RCA\_R\_PDA & $1.039 \times 10^{-7}$ & $1.178 \times 10^{11}$ & $1.228 \times 10^{4}$ & $3.70 \times 10^{-3}$ \\ \hline
Cap\_RCA\_R\_PDA2 & $7.532 \times 10^{-8}$ & $1.635 \times 10^{11}$ & $1.235 \times 10^{4}$ & $2.92 \times 10^{-3}$ \\ \hline
Cap\_RCA\_R\_PLB & $9.533 \times 10^{-8}$ & $1.281 \times 10^{11}$ & $1.226 \times 10^{4}$ & $3.78 \times 10^{-3}$ \\ \hline
Cap\_RCA\_R\_PLB2 & $5.995 \times 10^{-8}$ & $2.065 \times 10^{11}$ & $1.242 \times 10^{4}$ & $3.19 \times 10^{-3}$ \\ \hline
Cap\_RCA\_R\_PLB3 & $9.548 \times 10^{-8}$ & $1.205 \times 10^{11}$ & $1.175 \times 10^{4}$ & $2.08 \times 10^{-3}$ \\ \hline
Cap\_RCA\_RA1 & $1.081 \times 10^{-7}$ & $1.099 \times 10^{11}$ & $1.198 \times 10^{4}$ & $8.33 \times 10^{-3}$ \\ \hline
Cap\_Aorta\_Outlet & $3.496 \times 10^{-5}$ & $3.677 \times 10^{8}$ & $1.285 \times 10^{4}$ & $6.63 \times 10^{-3}$ \\ \hline
\end{tabular}
\caption{Flow rate and pressure at each Windkessel imposed boundary.}
\label{tab:table2}
\end{table}
\indent In addition to the visual observations, Table \ref{tab:table2} provides a quantitative comparison between the simulated outlet pressures and the calculated pressures derived from the relationship $P=Q\cdot R$. The results demonstrate a very small discrepancy. This close agreement highlights the accuracy and reliability of the implemented Windkessel boundary conditions. The minimal differences can be attributed to numerical approximations inherent in the simulation process but remain well within acceptable limits for hemodynamic simulations.\\
\indent It is important to note that, due to the incompressible flow assumption, only relative pressure holds physical significance. To ensure consistency, the resolved pressure values are referenced to a hardcoded baseline pressure of 8000 Pa in the current implementation. This approach preserves fidelity in the interpretation of the pressure field while adhering to the incompressible flow assumption.\\
\indent Overall, this case demonstrates the capability of the CoronaryHemodynamics framework to accurately simulate blood flow in physiologically realistic vascular geometries with complex boundary conditions. The results validate the robustness and reliability of the implemented Windkessel models, reinforcing their suitability for studies of coronary hemodynamics in anatomically accurate domains.
\begin{figure}[h]
\includegraphics[width=\textwidth]{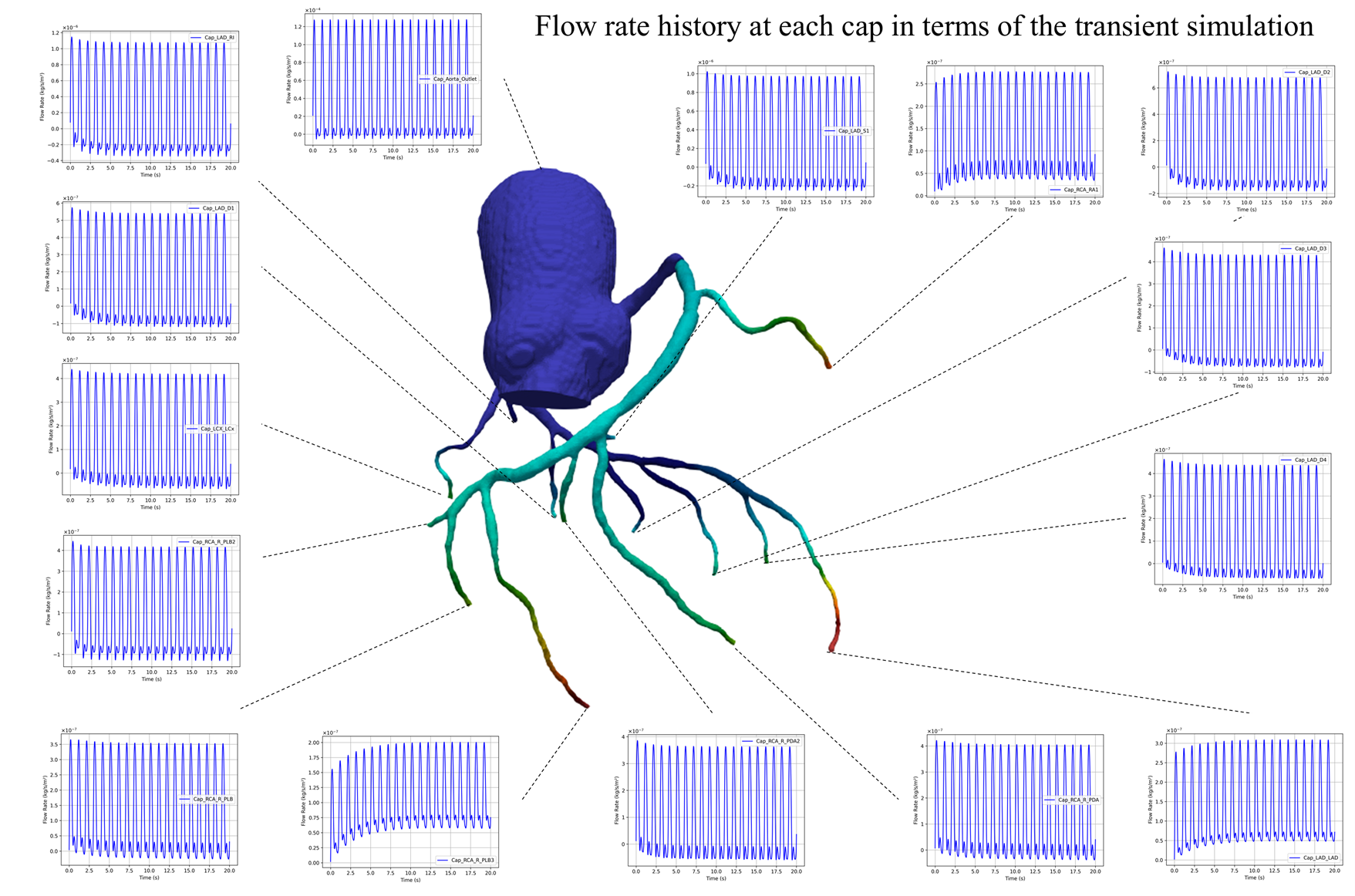}
\centering
\caption{Flow rate variation at each outlet cap across 10 pulse wave cycle}
\label{figs:fig11}
\end{figure}
\subsection{CoronaryFlow-transient simulation}

The second demonstration case highlights the capability of the CoronaryHemodynamics framework to simulate transient coronary blood flow dynamics over multiple cardiac cycles using realistic coronary artery geometry. This case emphasizes the solver’s ability to handle pulsatile flow conditions and accurately capture time-dependent variations in flow rate, pressure, and velocity fields. The simulations were performed using the transient solver \textit{pimpleCoronaryFoam}.
The mesh configuration for this case is identical to the previous steady-state simulation. The computational domain was evenly decomposed using the METIS decomposition method and executed on 8 cores of an AMD EPYC 7713 processor. The total computational time for the transient simulation was 24 hours, with a constant timestep of $ 1.0\times10^{-4} s$.\\
\indent Figure \ref{figs:fig11} presents the flow rate variations at each outlet cap across 10 pulse wave cycles. Initially, the flow rate exhibits oscillations as the simulation adjusts to the imposed pulsatile inlet conditions and outlet boundary constraints. Over time, the flow rate histories at each outlet stabilize into a periodic pattern consistent with the input waveform. This behavior demonstrates the solver's numerical stability and its capacity to simulate transient responses governed by the Windkessel boundary conditions.\\
\begin{figure}
\subfloat[9.0s]{\includegraphics[width = 0.2\linewidth]{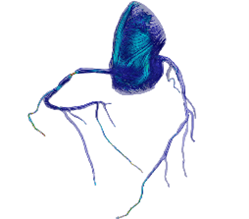}} 
\subfloat[9.1s]{\includegraphics[width = 0.2\linewidth]{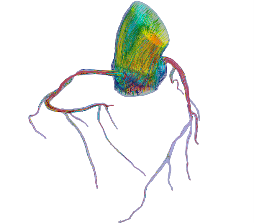}}
\subfloat[9.2s]{\includegraphics[width = 0.2\linewidth]{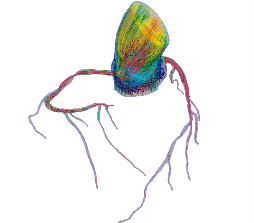}}
\subfloat[9.3s]{\includegraphics[width = 0.2\linewidth]{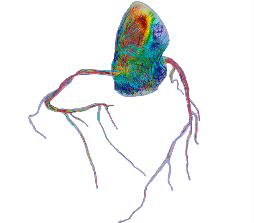}} 
\subfloat[9.4s]{\includegraphics[width = 0.2\linewidth]{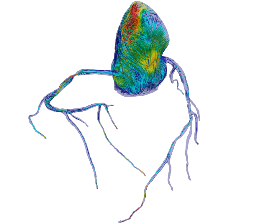}} \\
\subfloat[9.5s]{\includegraphics[width = 0.2\linewidth]{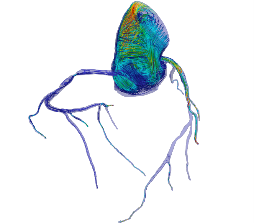}} 
\subfloat[9.6s]{\includegraphics[width = 0.2\linewidth]{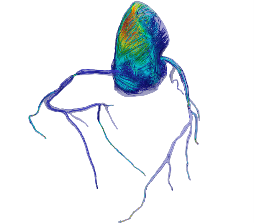}}
\subfloat[9.7s]{\includegraphics[width = 0.2\linewidth]{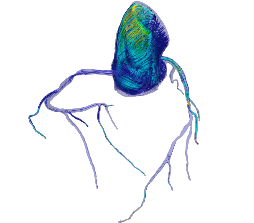}}
\subfloat[9.8s]{\includegraphics[width = 0.2\linewidth]{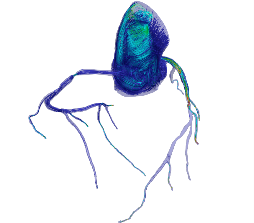}} 
\subfloat[9.9s]{\includegraphics[width = 0.2\linewidth]{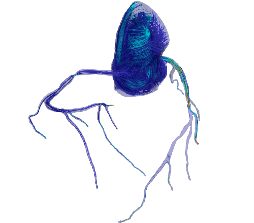}} \\
\caption{Velocity field color maps and streamlines across the 10th cycle of the pulse wave}
\label{figs:fig12}
\end{figure}
\indent To provide further insight into the flow dynamics, Figure \ref{figs:fig12} illustrates snapshots of the velocity field and streamlines at multiple time steps during the 10th cardiac cycle. These snapshots, taken at intervals of 0.1 seconds, reveal the evolution of the velocity field throughout the pulse wave cycle. The results show smooth and consistent flow patterns across the coronary network, capturing flow acceleration during systole and deceleration during diastole. This time-resolved analysis provides a detailed depiction of the transient flow behavior within the coronary arteries, including regions of high velocity and recirculation zones.\\
\indent The results of this demonstration emphasize the accuracy and robustness of the CoronaryHemodynamics package in handling complex, time-dependent hemodynamic scenarios. By successfully capturing the pulsatile nature of coronary blood flow, the solver demonstrates its suitability for studying physiological and pathological conditions that involve transient flow dynamics, such as cardiac cycle-dependent wall shear stress distribution and pressure variations.\\
\indent In summary, the second demonstration highlights the framework's ability to simulate realistic transient hemodynamics in anatomically accurate coronary artery geometries. The results further validate the reliability of the Windkessel boundary models in reproducing physiologically meaningful flow rate and pressure variations under pulsatile inlet conditions. This capability makes the CoronaryHemodynamics package a valuable tool for investigating time-dependent coronary flow behavior and related clinical applications.

\section{Discussions}
\label{diss}
The results presented in this study demonstrate the accuracy, robustness, and applicability of the CoronaryHemodynamics package in simulating coronary artery hemodynamics under both steady-state and transient conditions. The implemented Windkessel boundary models, integrated into the OpenFOAM-based framework, allow for physiologically realistic boundary conditions, which are critical for analyzing flow behaviors within complex vascular geometries.\\
\indent The validation results confirm the accurate implementation of both the two-element and three-element Windkessel models. In the pipe flow test cases, comparisons between the simulated and analytical solutions show excellent agreement for outlet pressure profiles. This minimal error highlights the reliability of the numerical solver and the robustness of the boundary condition implementation. Such close agreement reinforces the suitability of the package for solving complex flow dynamics where pressure and flow relationships play a dominant role.\\
\indent The steady-state simulation results illustrate the final distributions of wall shear stress pressure, and velocity across the coronary artery network. The stabilization of flow rate and pressure at each outlet to constant values aligns closely with theoretical predictions based on $P=Q\cdot R$. These observations validate the numerical solver’s ability to handle steady-state physiological conditions while maintaining consistency with the Windkessel boundary framework. Furthermore, the simulation results provide valuable insights into the hemodynamic environment within the coronary arteries. For instance, regions of elevated wall shear stress are identified, which are critical indicators for the development of atherosclerosis. Such detailed spatial distributions offer a deeper understanding of flow-induced stresses and their physiological implications.\\
\indent The transient simulation case further demonstrates the capability of the \textit{pimpleCoronaryFoam} solver to capture pulsatile flow dynamics over multiple cardiac cycles. The flow rate variations at the outlet boundaries stabilize into periodic waveforms consistent with the input conditions, indicating the numerical stability of the solver under time-varying boundary constraints. Additionally, the velocity field snapshots across the cardiac cycle reveal the detailed evolution of flow behavior, including acceleration during systole and deceleration during diastole. These results provide critical insights into transient flow dynamics, which are particularly relevant for understanding cardiac cycle-dependent phenomena such as pressure variations and wall shear stress distributions.\\
\indent The automation and streamlined workflow provided by the CoronaryHemodynamics package significantly reduce the manual workload typically associated with CFD simulations. The integration of tools like cfMesh for automatic meshing of complex anatomical geometries ensures consistent and high-quality meshes, which are crucial for numerical accuracy. The standardized naming convention and automated preprocessing steps further enhance the usability and reproducibility of the framework. The steady-state solver offers a computationally efficient alternative for cases where transient dynamics are not required. In contrast, the transient solver extends the package’s applicability to pulsatile flow conditions, capturing time-dependent phenomena critical for clinical and research applications.\\
\indent The demonstrated capabilities of the CoronaryHemodynamics package position it as a valuable tool for a wide range of applications. In cardiovascular research, the package can provide detailed analyses of wall shear stress and pressure distributions, which are key indicators for identifying regions susceptible to atherosclerosis. In clinical decision-making, it offers a means to simulate patient-specific coronary flow dynamics, supporting surgical planning and medical intervention evaluation. Furthermore, the ability to incorporate imaging-derived geometries highlights its potential to advance personalized medicine.\\
\indent While the results are promising, there are areas for future improvement. Model refinements could focus on enhancing the Windkessel models to account for non-linear arterial compliance and dynamic resistance. Additional validation with experimental data, such as in vivo or in vitro measurements, would further strengthen confidence in the results. Computational efficiency could be improved by optimizing the solvers for faster simulations, particularly for large-scale geometries or multi-cycle transient analyses. The inclusion of more physiological features, such as blood viscosity variations, non-Newtonian effects, or vascular wall elasticity, would further enhance the realism of the simulations.

\section{Declaration of Competing Interest}
\label{decl}
\indent The authors declare that they have no known competing financial interests or personal relationships that could have appeared to influence the work reported in this paper.

\section{Acknowledgments}
\label{ack}
\indent We would like to express our gratitude to the developers and contributors of BlueCFD-Core and OpenFOAM, whose open-source tools and frameworks provided the foundation for this work. Their innovative platforms have significantly advanced computational fluid dynamics, enabling robust and accessible simulations in diverse applications, including cardiovascular hemodynamics.

%% The Appendices part is started with the command \appendix;
%% appendix sections are then done as normal sections
%% \appendix

%% \section{}
%% \label{}

%% References
%%
%% Following citation commands can be used in the body text:
%% Usage of \cite is as follows:
%%   \cite{key}         ==>>  [#]
%%   \cite[chap. 2]{key} ==>> [#, chap. 2]
%%

%% References with bibTeX database:
\bibliographystyle{elsarticle-num}
\bibliography{bibtex}

\begin{thebibliography}{10}
\expandafter\ifx\csname url\endcsname\relax
  \def\url#1{\texttt{#1}}\fi
\expandafter\ifx\csname urlprefix\endcsname\relax\def\urlprefix{URL }\fi
\expandafter\ifx\csname href\endcsname\relax
  \def\href#1#2{#2} \def\path#1{#1}\fi

\bibitem{RN12}
T.~N. A.~M. Vuong, M.~Bartolf-Kopp, K.~Andelovic, T.~Jungst, N.~Farbehi, S.~G. Wise, C.~Hayward, M.~C. Stevens, J.~Rnjak-Kovacina, Integrating computational and biological hemodynamic approaches to improve modeling of atherosclerotic arteries, Advanced Science 11~(26) (2024) 2307627.
\newblock \href {https://doi.org/https://doi.org/10.1002/advs.202307627} {\path{doi:https://doi.org/10.1002/advs.202307627}}.

\bibitem{RN13}
M.~Roy, S.~Chakraborty, How does the stiffness of blood vessel walls and deposited plaques impact coronary artery diseases?, Physics of Fluids 36~(8) (2024).
\newblock \href {https://doi.org/10.1063/5.0226771} {\path{doi:10.1063/5.0226771}}.

\bibitem{RN14}
A.~Aminuddin, N.~Samah, U.~Vijakumaran, N.~A.~C. Roos, F.~M. Nor, W.~M. H.~W. Razali, S.~F. Mohamad, B.~B. Cong, F.~Hamzah, A.~Hamid, A.~Ugusman, Unveiling timps: A systematic review of their role as biomarkers in atherosclerosis and coronary artery disease, Diseases 12 (2024).
\newblock \href {https://doi.org/10.3390/diseases12080177} {\path{doi:10.3390/diseases12080177}}.

\bibitem{RN15}
G.~De~Nisco, C.~Chiastra, E.~Hartman, A.~Hoogendoorn, J.~Daemen, K.~Calò, D.~Gallo, U.~Morbiducci, J.~Wentzel, Comparison of swine and human computational hemodynamics models for the study of coronary atherosclerosis, Frontiers in Bioengineering and Biotechnology 9 (2021).
\newblock \href {https://doi.org/10.3389/fbioe.2021.731924} {\path{doi:10.3389/fbioe.2021.731924}}.

\bibitem{RN16}
M.~Biglarian, B.~Firoozabadi, M.~Saidi, Atheroprone sites of coronary artery bifurcation: Effect of heart motion on hemodynamics-dependent monocytes deposition, Computers in biology and medicine 133 (2021) 104411.
\newblock \href {https://doi.org/10.1016/j.compbiomed.2021.104411} {\path{doi:10.1016/j.compbiomed.2021.104411}}.

\bibitem{RN18}
B.~De~Bruyne, N.~Pijls, G.~Heyndrickx, D.~Hodeige, R.~Kirkeeide, K.~Gould, Pressure-derived fractional flow reserve to assess serial epicardial stenoses: theoretical basis and animal validation, Circulation 101 15 (2000) 1840--1847.
\newblock \href {https://doi.org/10.1161/01.CIR.101.15.1840} {\path{doi:10.1161/01.CIR.101.15.1840}}.

\bibitem{RN8}
W.~Guo, W.~He, Y.~Lu, J.~Yin, L.~Shen, S.~Yang, H.~Jin, X.~Wang, J.~Jun, X.~Hu, J.~Liang, W.~Wei, J.~Wu, H.~Zhang, H.~Zhou, Y.~Wu, R.~Yang, J.~Huang, G.~Tong, B.~Gao, R.~Chen, J.~Liu, Z.~Yan, Z.~Cheng, J.~Wang, C.~Li, Z.~Yao, M.~Zeng, J.~Ge, Ct-ffr by expanding coronary tree with newton–krylov–schwarz method to solve the governing equations of cfd, European Heart Journal - Imaging Methods and Practice 2~(3) (2024).
\newblock \href {https://doi.org/10.1093/ehjimp/qyae106} {\path{doi:10.1093/ehjimp/qyae106}}.

\bibitem{RN17}
J.~Jiang, C.-H. Du, Y.~Hu, H.~Yuan, J.~Wang, Y.~Pan, L.~Bao, L.~Dong, C.~Li, Y.~Sun, X.~Leng, J.~Xiang, L.~Tang, J.~Wang, Diagnostic performance of computational fluid dynamics (cfd)-based fractional flow reserve (ffr) derived from coronary computed tomographic angiography (ccta) for assessing functional severity of coronary lesions, Quantitative Imaging in Medicine and Surgery 13 (2023) 1672--1685.
\newblock \href {https://doi.org/10.21037/qims-22-521} {\path{doi:10.21037/qims-22-521}}.

\bibitem{RN9}
M.~Watarai, M.~Otsuka, K.~Yazaki, Y.~Inagaki, M.~Kahata, A.~Kumagai, K.~Inoue, H.~Koganei, K.~Enta, Y.~Ishii, Applicability of quantitative flow ratio for rapid evaluation of intermediate coronary stenosis: comparison with instantaneous wave-free ratio in clinical practice, The International Journal of Cardiovascular Imaging 35~(11) (2019) 1963--1969.
\newblock \href {https://doi.org/10.1007/s10554-019-01656-z} {\path{doi:10.1007/s10554-019-01656-z}}.

\bibitem{RN10}
S.~Smolka, A.~M. Fava, M.~Moshage, M.~Marwan, M.~Y. Desai, S.~Achenbach, Ct-ffr in predicting future cardiovascular events, European Heart Journal (2022).

\bibitem{RN11}
J.~Liu, B.~Mao, Y.~Feng, B.~Li, J.~Liu, Y.~Liu, Closed-loop geometric multi-scale heart-coronary artery model for the numerical calculation of fractional flow reserve, Computer methods and programs in biomedicine 208 (2021) 106266.

\bibitem{RN1}
A.~Updegrove, N.~M. Wilson, J.~Merkow, H.~Lan, A.~L. Marsden, S.~C. Shadden, Simvascular: An open source pipeline for cardiovascular simulation, Annals of Biomedical Engineering 45~(3) (2017) 525--541.
\newblock \href {https://doi.org/10.1007/s10439-016-1762-8} {\path{doi:10.1007/s10439-016-1762-8}}.

\bibitem{RN2}
C.~J. Arthurs, R.~Khlebnikov, A.~Melville, M.~Marcan, A.~Gómez, D.~Dillon-Murphy, F.~Cuomo, M.~S. Vieira, J.~Schollenberger, S.~R. Lynch, C.~Tossas-Betancourt, K.~Iyer, S.~Hopper, E.~Livingston, P.~Youssefi, A.~Noorani, S.~B. Ahmed, F.~J.~H. Nauta, T.~M. J.~v. Bakel, Y.~Ahmed, P.~A. J.~v. Bakel, J.~P. Mynard, P.~D. Achille, H.~Gharahi, K.~D. Lau, V.~Filonova, M.~Aguirre, N.~Nama, N.~Xiao, S.~Baek, K.~C. Garikipati, O.~Sahni, D.~A. Nordsletten, C.~A. Figueroa, Crimson: An open-source software framework for cardiovascular integrated modelling and simulation, PLoS Computational Biology 17 (2020).

\bibitem{RN3}
I.~Larrabide, P.~J. Blanco, S.~A. Urquiza, E.~A. Dari, M.~J. Vénere, N.~A. de~Souza~e Silva, R.~A. Feijóo, Hemolab – hemodynamics modelling laboratory: An application for modelling the human cardiovascular system, Computers in Biology and Medicine 42~(10) (2012) 993--1004.
\newblock \href {https://doi.org/https://doi.org/10.1016/j.compbiomed.2012.07.011} {\path{doi:https://doi.org/10.1016/j.compbiomed.2012.07.011}}.

\bibitem{RN4}
P.~C. Africa, I.~Fumagalli, M.~Bucelli, A.~Zingaro, M.~Fedele, A.~Quarteroni, lifex-cfd: An open-source computational fluid dynamics solver for cardiovascular applications, Computer Physics Communications 296 (2024) 109039.

\bibitem{RN37}
OpenFOAM, \href{https://openfoam.org/}{The openfoam foundation} (2024).
\newline\urlprefix\url{https://openfoam.org/}

\bibitem{RN38}
F.~D. Portugal, \href{https://bluecfd.github.io/Core/}{The bluecfd-core project} (2024).
\newline\urlprefix\url{https://bluecfd.github.io/Core/}

\bibitem{RN6}
C.~Fields, \href{https://cfmesh.com/cfmesh/}{Cf-mesh+} (2024).
\newline\urlprefix\url{https://cfmesh.com/cfmesh/}

\bibitem{RN5}
F.~Salmon, L.~Chatellier, pymeshfoam: Automated meshing for cfd and fluid–structure simulations, SoftwareX 23 (2023) 101431.
\newblock \href {https://doi.org/https://doi.org/10.1016/j.softx.2023.101431} {\path{doi:https://doi.org/10.1016/j.softx.2023.101431}}.

\bibitem{RN7}
N.~Westerhof, J.~W. Lankhaar, B.~E. Westerhof, The arterial windkessel, Med Biol Eng Comput 47~(2) (2009) 131--41.
\newblock \href {https://doi.org/10.1007/s11517-008-0359-2} {\path{doi:10.1007/s11517-008-0359-2}}.

\bibitem{RN21}
Z.~Li, W.~Mao, A fast approach to estimating windkessel model parameters for patient-specific multi-scale cfd simulations of aortic flow, Computers \& Fluids (2022).
\newblock \href {https://doi.org/10.1016/j.compfluid.2023.105894} {\path{doi:10.1016/j.compfluid.2023.105894}}.

\bibitem{RN25}
R.~Tricarico, S.~Berceli, R.~Tran‐Son‐Tay, Y.~He, Non-invasive estimation of the parameters of a three-element windkessel model of aortic arch arteries in patients undergoing thoracic endovascular aortic repair, Frontiers in Bioengineering and Biotechnology 11 (2023).
\newblock \href {https://doi.org/10.3389/fbioe.2023.1127855} {\path{doi:10.3389/fbioe.2023.1127855}}.

\bibitem{RN26}
R.~Spilker, C.~Taylor, Tuning multidomain hemodynamic simulations to match physiological measurements, Annals of Biomedical Engineering 38 (2010) 2635--2648.
\newblock \href {https://doi.org/10.1007/s10439-010-0011-9} {\path{doi:10.1007/s10439-010-0011-9}}.

\bibitem{RN20}
Z.~Li, W.~Mao, A fast approach to estimating windkessel model parameters for patient-specific multi-scale cfd simulations of aortic flow, Computers \& Fluids (2022).
\newblock \href {https://doi.org/10.1016/j.compfluid.2023.105894} {\path{doi:10.1016/j.compfluid.2023.105894}}.

\bibitem{RN22}
V.~Resmi, N.~Selvaganesan, Study on fractional order arterial windkessel model using optimization method, IETE Journal of Education 64 (2023) 103--111.
\newblock \href {https://doi.org/10.1080/09747338.2023.2210093} {\path{doi:10.1080/09747338.2023.2210093}}.

\bibitem{RN23}
Y.~Aboelkassem, Z.~Virag, A hybrid windkessel-womersley model for blood flow in arteries, Journal of theoretical biology 462 (2019) 499--513.
\newblock \href {https://doi.org/10.1016/j.jtbi.2018.12.005} {\path{doi:10.1016/j.jtbi.2018.12.005}}.

\bibitem{RN29}
B.~Bruyne, T.~Baudhuin, J.~Melin, N.~Pijls, S.~Sys, A.~Bol, W.~Paulus, G.~Heyndrickx, W.~Wijns, Coronary flow reserve calculated from pressure measurements in humans. validation with positron emission tomography, Circulation 89 3 (1994) 1013--1022.
\newblock \href {https://doi.org/10.1161/01.CIR.89.3.1013} {\path{doi:10.1161/01.CIR.89.3.1013}}.

\bibitem{RN28}
Y.~Huo, G.~Kassab, The scaling of blood flow resistance: from a single vessel to the entire distal tree, Biophysical journal 96 2 (2009) 339--346.
\newblock \href {https://doi.org/10.1016/j.bpj.2008.09.038} {\path{doi:10.1016/j.bpj.2008.09.038}}.

\bibitem{RN30}
Y.~Zhou, G.~Kassab, S.~Molloi, On the design of the coronary arterial tree: a generalization of murray's law, Physics in medicine and biology 44 12 (1999) 2929--2945.
\newblock \href {https://doi.org/10.1088/0031-9155/44/12/306} {\path{doi:10.1088/0031-9155/44/12/306}}.

\bibitem{RN32}
M.~Marcus, W.~Chilian, H.~Kanatsuka, K.~Dellsperger, C.~Eastham, K.~Lamping, Understanding the coronary circulation through studies at the microvascular level, Circulation 82 (1990) 1.
\newblock \href {https://doi.org/10.1161/01.CIR.82.1.1} {\path{doi:10.1161/01.CIR.82.1.1}}.

\bibitem{RN33}
J.~Bouwmeester, I.~Belenkie, N.~Shrive, J.~Tyberg, J.~Tyberg, Partitioning pulmonary vascular resistance using the reservoir-wave model, Journal of applied physiology 115 12 (2013) 1838--1845.
\newblock \href {https://doi.org/10.1152/japplphysiol.00750.2013} {\path{doi:10.1152/japplphysiol.00750.2013}}.

\bibitem{RN34}
P.~Virtanen, R.~Gommers, T.~E. Oliphant, M.~Haberland, T.~Reddy, D.~Cournapeau, E.~Burovski, P.~Peterson, W.~Weckesser, J.~Bright, S.~J. van~der Walt, M.~Brett, J.~Wilson, K.~J. Millman, N.~Mayorov, A.~R.~J. Nelson, E.~Jones, R.~Kern, E.~Larson, C.~J. Carey, I.~Polat, Y.~Feng, E.~W. Moore, J.~VanderPlas, D.~Laxalde, J.~Perktold, R.~Cimrman, I.~Henriksen, E.~A. Quintero, C.~R. Harris, A.~M. Archibald, A.~H. Ribeiro, F.~Pedregosa, P.~van Mulbregt, A.~Vijaykumar, A.~P. Bardelli, A.~Rothberg, A.~Hilboll, A.~Kloeckner, A.~Scopatz, A.~Lee, A.~Rokem, C.~N. Woods, C.~Fulton, C.~Masson, C.~Häggström, C.~Fitzgerald, D.~A. Nicholson, D.~R. Hagen, D.~V. Pasechnik, E.~Olivetti, E.~Martin, E.~Wieser, F.~Silva, F.~Lenders, F.~Wilhelm, G.~Young, G.~A. Price, G.-L. Ingold, G.~E. Allen, G.~R. Lee, H.~Audren, I.~Probst, J.~P. Dietrich, J.~Silterra, J.~T. Webber, J.~Slavič, J.~Nothman, J.~Buchner, J.~Kulick, J.~L. Schönberger, J.~V. de~Miranda~Cardoso, J.~Reimer, J.~Harrington, J.~L.~C. Rodríguez, J.~Nunez-Iglesias,
  J.~Kuczynski, K.~Tritz, M.~Thoma, M.~Newville, M.~Kümmerer, M.~Bolingbroke, M.~Tartre, M.~Pak, N.~J. Smith, N.~Nowaczyk, N.~Shebanov, O.~Pavlyk, P.~A. Brodtkorb, P.~Lee, R.~T. McGibbon, R.~Feldbauer, S.~Lewis, S.~Tygier, S.~Sievert, S.~Vigna, S.~Peterson, S.~More, T.~Pudlik, T.~Oshima, et~al., Scipy 1.0: fundamental algorithms for scientific computing in python, Nature Methods 17~(3) (2020) 261--272.
\newblock \href {https://doi.org/10.1038/s41592-019-0686-2} {\path{doi:10.1038/s41592-019-0686-2}}.

\bibitem{RN35}
Q.~Li, Y.~Zhang, C.~Wang, S.~Dong, Y.~Mao, Y.~Tang, Y.~Zeng, Diagnostic performance of ct-derived resting distal to aortic pressure ratio (resting pd/pa) vs. ct-derived fractional flow reserve (ct-ffr) in coronary lesion severity assessment, Ann Transl Med 9~(17) (2021) 1390.
\newblock \href {https://doi.org/10.21037/atm-21-4325} {\path{doi:10.21037/atm-21-4325}}.

\bibitem{RN36}
D.~Fu, M.~Liu, M.~Shao, Y.~Mao, C.~Li, H.~Jiang, X.~Li, Functional evaluation of percutaneous coronary intervention based on ct images of three‐dimensional reconstructed coronary artery model, Contrast Media \& Molecular Imaging 2023~(1) (2023) 6761830.

\end{thebibliography}

%% Authors are advised to submit their bibtex database files. They are
%% requested to list a bibtex style file in the manuscript if they do
%% not want to use elsarticle-num.bst.

%% References without bibTeX database:

% \begin{thebibliography}{00}

%% \bibitem must have the following form:
%%   \bibitem{key}...
%%

% \bibitem{}

% \end{thebibliography}

\end{document}